\documentclass[english]{cccconf}
\usepackage[comma,numbers,square,sort&compress]{natbib}
\usepackage{epstopdf}
\usepackage{amssymb}
\usepackage{amsmath}
\usepackage[colorlinks,
            linkcolor=blue,
            anchorcolor=blue,
            citecolor=blue
            ]{hyperref}

\begin{document}

\title{  Quantum-Inspired Solvers on Mixed-Integer Linear Programming Problem}

\author{Hao Wang\aref{SCUT},
        Yu Pan\aref{ZJU},
        Wei Cui\aref{SCUT}
    }


\affiliation[SCUT]{School of Automation Science and Engineering, South China University of Technology, Guangzhou~510641, China \email{aucuiwei@scut.edu.cn}
}
\affiliation[ZJU]{College of Control Science and Engineering, Zhejiang University,
Hangzhou 310058, China
}

\maketitle

\begin{abstract}
Mixed-integer linear programming (MILP) plays a crucial role in artificial intelligence, biochemistry, finance, cryptography, etc. Notwithstanding popular for decades, the researches of MILP solvers are still limited by the resource consumption caused by complexity and failure of Moore's Law. Quantum-inspired Ising machines, as a new computing paradigm, can be used to solve integer programming problems by reducing them into Ising models. Therefore, it is necessary to understand the technical evolution of quantum inspired solvers to break the bottleneck. In this paper, the concept and traditional algorithms for MILP are introduced. Then, focused on Ising model, the principle and implementations of annealers and coherent Ising machines are summarized. Finally, the paper discusses the challenges and opportunities of miniaturized solvers in the future.
\end{abstract}


\footnotetext{This work was supported by the National Key R\& D Program of China under Grant 2018YFB1700100, the National Natural Science
Foundation of China under Grant 61873317.}

\section{Introduction}

The demand of solvers on NP-hard problems is ubquitous in daily life, and it has attracted tremendous attention from both academia and industry recently. Many combinatorial optimization problems are NP-hard problems, e.g., knapsack problems\cite{Martello1990kp} and traveling salesman problems\cite{Miller1960tsp}. At present, there is no polynomial time solution for this kind of problems. Solving large-scale NP-hard problems is notoriously difficult even for preeminent mathematicians or poweful supercomputers. Nevertheless, seeing the enormous impact it takes, such as the reduction of resource consumption, promotion of benefit, and even survival of enterprise, it is crucially important of seeking for efficient solutions to such puzzles. Karp described 21 NP-complete problems by showing the polynomial time many-one reduction from the boolean satisfiability problem\cite{Karp1972npc}. Actually, once one of these NP-complete problems is polynomial solvable, so are the others in principle.

One of the most vital NP-hard problems is mixed-integer linear programming (MILP) problems\cite{Wolsey2020ip}. The discrete-continuous optimization scheme is center in many applications including finance portfolio\cite{Jorion1992po}, vehicle routing\cite{Achuthan1991vrp}, and drug molecule design\cite{Allouche2014pd}. In the interest of solving the problem, a variety of commercial or open source solvers have been spawned on the basis of both precise algorithms such as branch and bound\cite{Land2010bnb} and approximate algorithms. Known as the first commercial MILP solver, Xpress was originally developed in 1986\cite{Ashford2007xpress}. Subsequently, Gurobi\cite{gurobi}, CPLEX\cite{cplex} and other commercial solvers have mushroomed over the past 30 years, as shown in the table\ref{tab:csolver}. SCIP\cite{scip}, one of the fastest non-commercial solvers for MILP problems, was originally developed in 2001, and has inspired the design of various solvers in our country. Notwithstanding the thriving solvers, with the problem scaling, there is still an uncoordinated contradiction between resource consumption and solution precision, which will be no less disastrous than the developing bottleneck of classical computers within saturation limits of Moore's Law.

\begin{table}[!htb]
\centering
\caption{Information about MILP solvers}
\label{tab:csolver}
\begin{tabular}{c|c|c|c}
        \hhline
        Solver & Country & Year & commercial \\ \hline
        Xpress & Britain & 1986 & Y \\ \hline
        CPLEX & America & 1988 & Y \\ \hline
        GLPK & Russia & 2000 & Y \\ \hline
        SCIP & Germany & 2001 & N \\ \hline
        Ipsolve & Finland & 2004 & Y \\ \hline
        Gurobi & America & 2009 & Y \\ \hline
        CMIP & China & 2018 & N \\ \hline
        COPT & China & 2019 & Y \\
        \hhline
\end{tabular}
\end{table}

Quantum computing is promoted as one of the potential ways to efficiently solve NP-hard problems and overcome the bottleneck. In recent years, many quantum algorithms have erupted and showed their novelty and potential in solving NP-hard problems, such as QAOA\cite{Farhi2014qaoa} and VQE\cite{Kandala2017vqe}. Some effort on quantum computing has also focused on hard optimization problems\cite{Qi2020cqn, Qi2021oqn, Yan2020nqn, Mu2020rlr}. Whereas, some non-von Neumann computers, such as general quantum computers and quantum-inspired Ising machines, have shown quantum advantages in max-cut problems\cite{Farhi2014qaoa}, protein folding problems\cite{Robert2021pf} and so on. Quantum-inspired solvers for various optimization problems have been springing up, especially quantum annealers based on adiabatic quantum computation\cite{Das2008qa}, and coherent Ising machines based on optical oscillators\cite{Takata2012laser, Wang2013dopo}. Generally, these solvers reduces a practical issue into the Ising model or quadratic unconstrained binary opitimization problem at first, and then embeds the problem in the corresponding hardware. The solution is measured once the machine has stopped evoluting.

\begin{figure*}[!htb]
        \centering
        \includegraphics[width=\hsize]{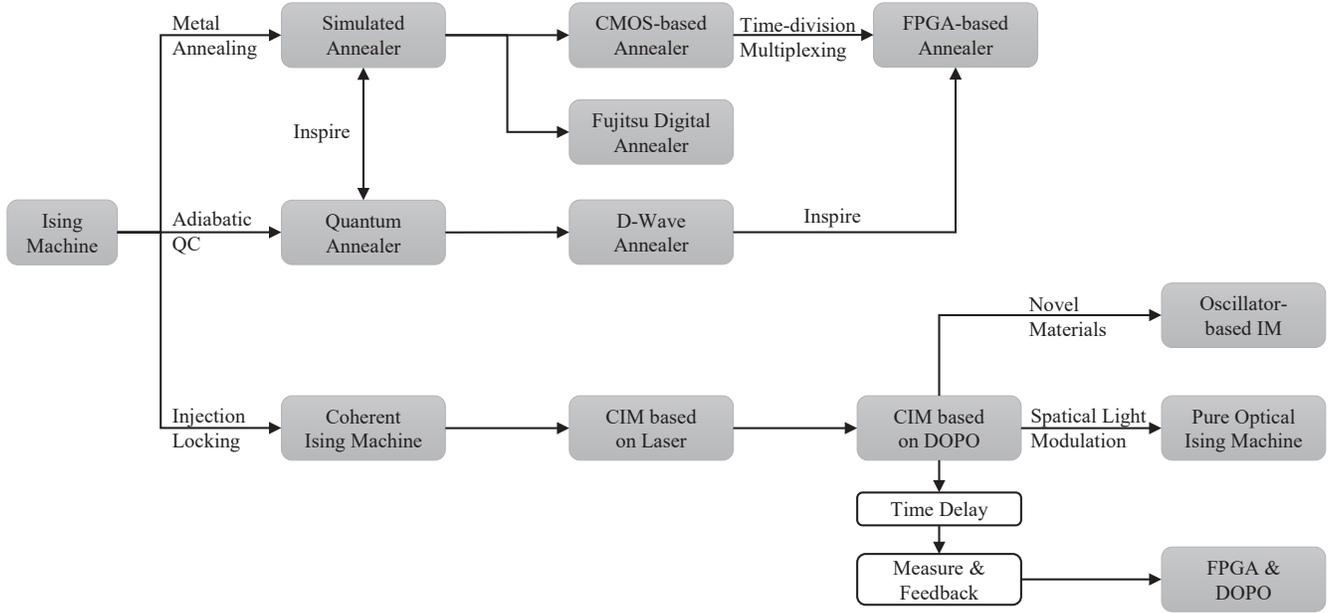}
        \caption{The evolution process of Ising machines.}
        \label{fig:evolution}
\end{figure*}

Due to the parallelism caused by quantum superposition and the tunneling effect caused by quantum fluctuation, quantum solvers have great advantages over the classical solvers in precision or efficiency. However, these quantum computers either required rigorous physical conditions such as ultra-low temperatures and confined spaces, or possessed huge volume, facing tremendous limitations to practical applications. In this paper, we introduce the working principle of several new type of computers represented by Ising machine, of which the main evolutionary process is presented in Figure~\ref{fig:evolution}.

The rest of this paper is organized as follows. In Section~\ref{sec:milp} we introduce the MILP Problem and some classical algorithms used for solvers. We discuss two main kinds of Ising machines in Section~\ref{sec:ising}, and end by analyzing the challenges and opportunities of quantum-inspired solvers. In Section~\ref{sec:conclude}, we conclude the paper.

\section{Mixed-Integer Programming\label{sec:milp}}

As a special case of linear programming (LP) problem, mixed-integer linear programming is defined as:
\begin{equation}
\label{eq:milp}
\begin{aligned}
        \min_{x}~&c^Tx \\
        s.t.~&Ax\leq b \\
        &x\in \mathbb{Z}^m \times \mathbb{R}^{n-m},
\end{aligned}
\end{equation}
where $c\in \mathbb{R}^n$ and $b \in \mathbb{R}^r$ are two coefficient vectors, $A \in \mathbb{R}^{r\times n}$ is a coefficient matrix. Equation~\ref{eq:milp} tells us that notwithstanding the discrete-continuous variables, the constraints and objective functions of MILP are tantamount to those of LP.

Regardless of efficiency, LP problems can be solved by methods such as simplex method, interior point method, ellipsoid method\cite{Wolsey2020ip}. Different from LP, it is the discrete variables of MILP, that is, the discontinuity of feasible region, that makes the above methods cannot be directly applied into MILP.

A trivial method was enumeration, which searched all feasible values of discrete variables inefficiently. A modified implicit enumeration method worked in pure integer programming problems that could be reduced into binary integer linear programming (BILP) problems\cite{Wolsey2020ip}. However, the aforementioned method is expensive even for pure integer programming, meaning that it is necessary to seek more efficient algorithms. Aiming at the inherent nature of MILP, methods including branch and bound, cutting plane, and column generation have been proposed\cite{Land2010bnb, Kelley1960cp, Lub2005dwcg}. In addition, decomposition algorithms such as DW decomposition\cite{Lub2005dwcg} are worthy for reference as well.

In some practical applications, expensive resource and time consumption seriously obstruct the precise computing for the minimum, motivating many heuristic and approximate algorithms. A computational cheap heuristic method for MILP is accepting the rounding of corresponding LP solutions. Though the rounding solution is not even a feasible one, it has heuristiced many algorithms with supplementary constraints, such as local branching\cite{Fischetti2008lb}.

Besides, algorithms including hill-climbing\cite{Hinson1983hill} have been inspired by a core idea named local search. The local search scheme started from one or several initial solutions, and ended with the optimal solution in certain feasible region by iterating searching for candidate solutions in the solution domain. Approximate algorithms inspired by biological or physical phenomena have been also used to design solvers, such as simulated annealing\cite{Laarh1987sa}. Almost all commercial and non-commercial solvers support both precise and approximate algorithms.

\section{Quantum-inspired Solvers\label{sec:ising}}

\subsection{Ising Model and Reduced MILP problems}

Many combinatorial optimization problems can be reduced into Ising problems\cite{Lucas2014Ising}. Ising model is a general mathematical model in statistical physics, which was firstly used to explain the correlation between spin states in ferromagnetic material and the macroscopic magnetic moment\cite{Ising1925}. It assumes that ferromagnetic matter consists of a collection of small magnetic needles, of which each is only allowed to be either $+1$ or $-1$ (in quantum mechanics it is described as $\left|\uparrow\right>$ and $\left|\downarrow\right>$). Adjacent magnetic needles interact with each other through field, combined with stochastic magnetic flips caused by the interference of ambient thermal noise.

The purpose of Ising problem is to minimize the Ising Hamiltonian, which is given by the following equation:
\begin{equation}
\label{eq:isingspin}
 H=-\sum_{i,j}^{V}J_{ij}\sigma_i\sigma_j-\sum_i^Vh_i\sigma_i,
\end{equation}
where $V$ represents the number of spins, $J_{ij}$ means the coupling constant between spins, the sum of $h_i\sigma_i$ is called external field, and $\sigma_i$ denotes spins. Obviously, each different combination of spin states corresponds to a unique value of energy. The whole system tends to have a lowest energy, corresponding the minimum loss function pursuing for.

Ising machine can solve many NP-complete problems, including BILP problems. As for BILP problem, for example, the Ising Hamiltonian could be divided into two parts\cite{Lucas2014Ising}:
\begin{equation}
\label{eq:ha}
\begin{aligned}
        H_A&=A\sum_{j=1}^m{{\left[b_j-\sum_{i=1}^N{S_{ji}x_i}\right]}^2}\\
        H_B&=-B\sum_{i=1}^N{c_ix_i},
\end{aligned}
\end{equation}
where $0<B\ll A$ are two positive integers, of which the value is chosen subjectively but bounded by the properties of constraint equations and the cost function. $x_i$ denotes a binary variable. $b_i$, $c_i$, $S_{ji}$ corresponds to the coefficients of the constraint $Sx=b$ and the cost $c^Tx$.

BILP is directly reduced to the Ising problems by Equation~\ref{eq:ha}, solved by the general Ising machine. However, for large scale MILP problems, the reduction from a discrete-continuous puzzle to a discrete conbinatorial optimizatiton problem is impractical. Fortunately, such puzzles have been able to be partitioned into solvable Ising problems combined with classical efforts\cite{Ajagekar2020dco}.

\subsection{Annealing-based Ising machine}

\begin{figure}[!htb]
        \centering
        \includegraphics[width=\hsize]{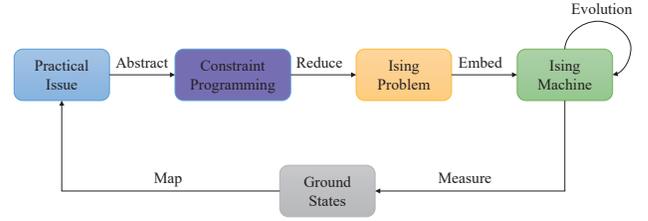}
        \caption{The diagram for solving a practical issue on an Ising machine.}
        \label{fig:solver}
\end{figure}

Ising machine is a solver used to solve Ising problems, that is, a physical machine that minimizes the Ising Hamiltonian in Equation~\ref{eq:isingspin}. Ising machine belongs to natural computing, a novel non-von Neumann computing paradigm in which the physical nature phenomenon used for calculation has an intrinsic convergence property\cite{Yamaoka2015cmos}. As shown in Figure~\ref{fig:solver}, to solve a practical issue, the puzzle should be mapped to the natural physical phenomenon corresponding to the convergence property. The solution to the puzzle is measured when the system stopped converging under the external field and evolution.

The most famous Ising machine suppose to be quantum annealer based on adiabatic quantum computation. Compared with classical simulated annealer, quantum annealing is superior in approximate optimization problems due to superposition and tunneling effect. Quantum annealers produced by companies such as D-Wave have been operational for commercial use\cite{Johnson2011dwave, Mc2020tvpdwave}.

While real quantum annealers are limited by the harsh operating environment, analog ones inspired by quantum annealing have demonstrated their superiority in some practical applications. The Digital Annealer developed by Fujitsu simulated the quantum annealing process by Monte Carlo method, demonstraing its power on solving complex combinatorial optimization problems\cite{Aramon2019fujitsu}. Benchmark test on real problems has demonstrated the strengths and weaknesses of different quantum and analog annealers\cite{Oshiyama2022benchmark}. However, these special computers are too cumbersome so that it is inconvenient for a global rollout.

Yamaoka et al.\cite{Yamaoka2015cmos} has proposed to use CMOS circuits for annealing, with the Ising model encoded in SRAM arrays. The coupling between spins is simulated by logic circuits, and the machine is able to solve MAX-CUT problems no more than 20k spins. They then have proposed a modified annealing processor based on FPGA\cite{Yoshimura2016fpga}. By minor embedding and time-division multiplexing architecture, the FPGA Ising machine has been designed to address large scale combinatorial optimization problems\cite{KYamamoto2019fpga} with almost any topology. The basic idea of FPGA-based annealer is to embed the Ising model into the electronic circuit, with each Ising spin mimicked by a circuit diagram equiping a memory cell array. In such a spin circuit unit, different coefficients are connected with coupling by AND gates, followed by an operator circuit controlled by some random pulses. The proposed FPGA annealers achieve the minimum value by local searching, i.e. flipping one bit and find the lowest energy repeatedly. To escape from the local optimum, probabilistic behavior such as pseudorandom number generator who generates aforementioned random pulses is added to the circuit to invert the spin. Besides, though it has been only used for SAT problem and integer factorization with several fundamental circuit structures, the invertible probabilistic circuit implemented by FPGA is also an alternative for such a global optimum without the necessity of complex inverter structure\cite{Aadit2021sim}.

\begin{figure}[!htb]
        \centering
        \includegraphics[width=\hsize]{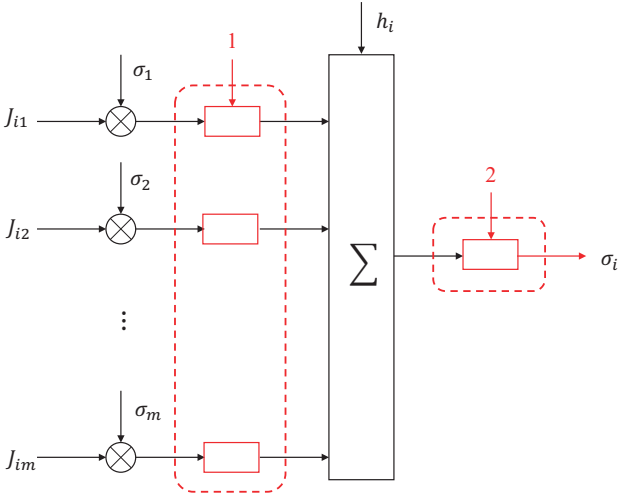}
        \caption{An implementation of Ising spin unit in the FPGA solver. The Monte Carlo method inspired by Quantum Annealing process is reserved in two interfaces that have been noted by dotted lines. Interface $1$ is used to invert input spins randomly, while interface $2$ is used to invert the output of the spin unit stochastically.}
        \label{fig:fpga}
\end{figure}

A feasible portable solver designed in FPGA as mentioned before is shown in Figure~\ref{fig:fpga}. A quantum-inspired annealer in FPGA votes possible candidates to get over the local minimum. The core to simulate quantum annealer on a classical FPGA is to update spins by appropriate Monte Carlo such as Gibbs sampling. For the Ising Hamiltonian in Equation~\ref{eq:isingspin}, a most common sampler requires an interaction of:
\begin{equation}
\label{eq:interaction}
 I_i=-\frac{\partial H}{\partial \sigma_i}=\sum_{j}^{V}J_{ij}\sigma_j+h_i.
\end{equation}
$I_i$ is used to vote candidate spins with corresponding Boltzmann probability:
\begin{equation}
\label{eq:boltzmann}
p_i \propto \exp{\left[-\frac{H(\sigma)}{kT}\right]}.
\end{equation}
The update unit for a single spin is implemented by the unit in Figure~\ref{fig:fpga}, consisting of Equation~\ref{eq:interaction} with corresponding probability in Equation~\ref{eq:boltzmann}. In FPGA, Equation~\ref{eq:interaction} is calculated by circuits, and the accept-reject probability in Equation~\ref{eq:boltzmann} is generated by randomly inverting the spins through Monte Carlo methods to get over the local minimum. Different units update spins by aforementioned interaction methods iteratively. Finally, the FPGA solver votes the ground state of corresponding Ising Hamiltonian, which is also the the solution to the question.

In some situations, it is ridiculous to directly take an expensive solver to solve an easy exercise or a puzzle undetermined whether unsolvable or not. Since FPGA-based annealers encode a problem with circuits, it is natural to simulate a proposed solver on CPU, GPU or TPU\cite{Aadit2021sim}. A mature simulated solver will be implemented even on PC, convenient for daily issues of small scale.

\subsection{Oscillator-based Coherent Ising Machine}

Coherent Ising machine (CIM) represents Ising spins by different polarization states above the oscillation threshold in coupled nonlinear oscillators, such as $0$-phase and $\pi$-phase. The coupling mechanism is based on injection locking, a common phenomenon in most nonlinear oscillators where the output frequency is precisely locked to a similar given external interference frequency\cite{Wang2017oim}. Injection locking consists of fundamental frequency locking or harmonic locking. Sub-harmonic injection locking (SHIL) is popular in CIM since it excites multiple stable phase-locked responses in oscillators.

The energy landscape of the initial Ising Hamiltonian is embedded into a network of oscillators by coupling mechanism. The ground states of Ising Hamiltonian, also corresponding to the solution of the original problem, is found through phase transitions of oscillations caused by external distractions. Differing from quantum and analog annealing processors, CIM gradually raises the gain to search for the minimum value of energy landscape upward, while annealers optimize along the landscape\cite{Marandi2014tmcim}.

Early researches on CIM began with the concept of an injection synchronous laser Ising machine proposed by Stanford in 2011\cite{Utsunomiya2011laser, Takata2012laser}. The number of lasers corresponds to the number of qubits but the coupling is difficult. They then have proposed a modified CIM based on degenerate optical parametric oscillators (DOPO) network\cite{Wang2013dopo}. Each degenerate OPO in the network denotes an Ising spin. All OPOs are in squeezed vacuum state under the threshold, while the phases of OPOs polarize into binary phase states above threshold by external pump. The coupling comes from mutual injection between different OPOs. Since strong coupling strength is able to cause probable errors in solving Ising problems, it is essential to choose a suitable coupling method. Time delay has been a potential method to generate suitable couplings, which modulates arbitrary Ising problems with $n$ spins by delaying $n-1$ times\cite{Marandi2014tmcim}.

DOPO-based CIM has been used to solve many problems\cite{Takata2015qcdopo, Takata2016cim16, Inagaki2016cim10k}. However, the scalability of CIM is poor. On the one hand, decoherence and dispersion effects of photons extremely limit the number of couplings in physical Ising machines. On the other hand, precise control of thousands of delay lines has remained impractical. They then have realized the coupling process of time multiplexed DOPO pulses with a measurement and feedback scheme implemented by FPGA\cite{Haribara2015qmfc, Haribara2016maxcut, Inagaki2016cim2k, Honjo2021cim10k}. Pierangeli et al. abandoned traditional mechanical networks and built a new Ising machine using spatial light modulation\cite{Pierangeli2019spatial}, breaking the scale limitation of optical Ising machine. Novel materials such as nanomagnet and nanophotonic also have been supposed to be the candidate for Ising model\cite{Sutton2017nanomag, Wang2018nanophoto}.

In fact, optical material is not the necessary condition for CIM, although photonic computing have been demonstrated more efficient in Boolean computation than von Neumann computers. As aforementioned, injection locking is universal in nonlinear oscillators, that is, almost all nonlinear oscillators are able to denote Ising spins. For example, LC, MEMS, and biochemical reaction networks have been suggested to be suitable as well as Photonics oscillators\cite{Wang2017oim, Chou2018wim}.

\subsection{Discussion on Miniaturized Solvers}

Aforementioned Ising machines have their own advantages and disadvantages on solving practical MILP problems. A qualified solver should be convenient for practical usage, that is, it is supposed to be miniaturized without harsh environment. Therefore, the following topics are needed to study further.

\begin{itemize}
  \item[(1)] \textbf{Scalabilty of annealers}. \\
Annealers have worked well in many practical applications, but the challenge is how to scale up the problem size without excessive resource consumption. Though minor embedding method is helpful as well as Chimera graph, it still stands a limitation of sizes. The time-division multiplexing architecture seems to be helpful, but more actual experiments are needed. For many large complex problems, the process of finding an optimal embedding graph is itself NP-hard, restricting the generalization and miniaturization vastly. Heuristic algorithms might help find an efficient embedding\cite{Cai2014ha}, but at the expense of reducing acuracy. Hence, it is significant to find more efficient embedding algorithm to scale up and generalize annealers.

  \item[(2)] \textbf{Limitation of CIMs from materials}. \\
Photonics computers requires sufficient space to generate optical pulses, which is the base of coupling. Methods such as time-multiplexing and measurement and feedback couldn't do anything to miniaturization of optical generation. Ising machine based on LC oscillators seems suitable for miniaturized solvers, but facing the same limitation of scalability. As for new nonlinear oscillators, how to couple spins is still an obstacle as well as the coupling performance.

  \item[(3)] \textbf{The reduction of MILP problems}. \\
BILP scheme can be reduced into Ising model directly. However, out of BILP, other MILP problems can't be embedded into Ising machine straightforwardly since Ising model is not capable for continuous variables. Although it is operable to split or convert MILP problems into BILP problems, more efforts have to be made and the quantum superior is then diminished. A potential direction is to reduce the MILP problem into XY model, whose Hamiltonian is given as
$$H = -\sum_{i,j}^VJ_{ij}\cos(\theta_i-\theta_j)-\sum_{i}^Vh_i\cos\theta_i.$$ The MILP problem then can be converted into finding the minimum value of the mixture of XY Hamiltonian and Ising Hamiltonian. Nevertheless, the solvers on XY model such as Born machines is still under theory\cite{Gomez2021born}. For further studies, the reduction of MILP will be annoying.

  \item[(4)] \textbf{More possible solvers}. \\
An alternative for physical solver is simulating software. The work principle of FPGA and LC oscillators can be simulated on classical computers, meaning it is possible to design and implement a virtual Ising machine on classical computer. The evolution can be speeded up by CPU, GPU, and TPU. Another miniaturization paradigm is the Nuclear Magnetic Resonance (NMR) quantum computer, which uses particles in resonance excitation in a magnetic field to produce excited state $\left|1\right>$ and ground state $\left|0\right>$ respectively, solving MILP problems in general quantum computing paradigm. General quantum computers, such as superconducting computer and optical computer, survives on a harsh physical environment which is available only in the laboratory now, while NMR computers can run fine at room temperature. The prototype of desktop quantum computer has been realized by SpinQ, though with a size no more than 3-bits\cite{Hou2021spind, Feng2022spinq3}. Obviously, it still needs more efforts to be used in practical applications.
\end{itemize}

\section{Conclusion\label{sec:conclude}}

In this paper, we have reviewed some quantum-inspired solvers from the perspective of working mechanism. Annealers based on adiabatic quantum computation find the mimnimum energy by searching along the landscape and speed up by quantum tunneling effect. Coherent Ising machines based on oscillators excite oscillators to find the mimnimum value upward. Analog solvers inspired by these two kinds of Ising machines are capable to solve Ising problems more efficiently than classical solvers. These solvers demonstrate their advantages and disadvantages on solving practical problems. However, there still exist many obstacles to design a miniaturized solver on MILP problems. We have listed many worthy directions above. In the future, more efficient solvers will be widely used to solve general MILP problems.

\end{document}